\documentclass[aps,prx,twocolumn,floatfix, showpacs,longbibliography]{revtex4-1}
\usepackage{graphicx}
\usepackage{multirow}
\usepackage{natbib}
\usepackage{comment}
\usepackage{color} 
\usepackage[colorlinks=true]{hyperref}

\begin{document}


\title{Creating Two-Dimensional Electron Gas in Nonpolar Oxide Interface via Polarization Discontinuity: First-Principles Analysis of  CaZrO$_3$/SrTiO$_3$ Heterostructure}

\author{Safdar Nazir, Jianli Cheng, and Kesong Yang}
\email{kesong@ucsd.edu,+1-858-534-2514}
\affiliation{Department of NanoEngineering, University of California, San Diego, 9500 Gilman Drive, Mail Code 0448, La Jolla, CA 92093-0448, USA}

\begin{abstract}

We studied strain-induced polarization and resulting conductivity in the nonpolar/nonpolar CaZrO$_3$/SrTiO$_3$ (CZO/STO) heterostructure (HS) system by means of first-principles electronic structure calculations. By modeling four types of CZO/STO HS-based slab systems, \textit{i.e.}, TiO$_2$/CaO and SrO/ZrO$_2$ interface models with CaO and ZrO$_2$ surface terminations in each model separately, we found that the lattice-mismatch-induced compressive strain leads to a  strong polarization in the CZO film, and as the CZO film thickness increases, there exist an insulator-to-metal transition. The polarization direction and critical thickness of the CZO film for forming interfacial metallic states depend on the surface termination of CZO film in both types of interface models. In the TiO$_2$/CaO and SrO/ZrO$_2$ interface models with CaO surface termination, the strong polarization drives the charge transfer from the CZO film to the first few TiO$_2$ layers in the STO substrate, leading to the formation of Two-Dimensional Electron Gas (2DEG) at the interface. 
In the HS models with ZrO$_2$ surface termination, two polarization domains with opposite directions were formed in the CZO film, which results in the charge transfer from the middle CZO layer to the interface and surface, respectively, leading to the co-existence of the 2DEG on the interface and the Two-Dimensional Hole Gas (2DHG) at the middle CZO layer. These findings open a new avenue to achieve 2DEG (2DHG) in perovskite-based HS systems via polarization discontinuity.

\end{abstract}

\pacs{73.20.-r, 73.40.-c, 71.15.-m}

\maketitle    

\section{Introduction}
The perovskite-based oxide heterostructures (HS) are attracting increasing interests because of their novel interfacial properties such as the interfacial superconductivity and ferromagnetism that are drastically different from those of the corresponding bulk materials.\cite{Brinkman_2007_NM,Bert_2011_NP,Richter_2011_NP,Bi_NC_2014} One typical example is the formation of the high-mobility Two-Dimensional Electron Gases (2DEG) at TiO$_2$-terminated interface in the polar/nonpolar LaAlO$_3$/SrTiO$_3$(LAO/STO) HS system.\cite{Ohtomo_2004_N,Okamoto} 
The excellent conducting property of the 2DEG  at the interface of the HS makes it promising for applications in the next-generation nanoelectronic oxide devices.\cite{Kalisky_2013_NM, Mannhart_2010_Sc,Sulpizio_ARMR_2014,Stemmer_ARMR_2014} Generally speaking,  there are three main approaches to form a 2DEG in a HS system. The first one is via modulation doping,\cite{Dingle_APL_1978} which often creates a 2DEG in a HS that consists of a wide-band-gap and a narrow-band-gap semiconductor. 
The \textit{n}-type doping in the wide-band-gap semiconductors introduces additional electrons in the HS, and the electrons transfer from the high conduction band edge of the wide-band-gap semiconductor to the conduction band of the narrow-band-gap semiconductor on the other side of the HS leads to the 2DEG.\cite{Stemmer_ARMR_2014} 
One example of the 2DEG utilizing this mechanism is the band bending AlGaAs/GaAs HS system, where the electrons forming the 2DEG in the interfacial GaAs layer are supplied from the doped AlGaAs.\cite{Hiyamizu_APL_1980,Harris_JAP_1987} The second one is by means of electronic reconstruction caused by the polar discontinuity at the interface between a polar and a nonpolar layer in a HS system such as the LAO/STO. In the LAO/STO HS system, the charge transferred from the polar (LaO)$^{1+}$ layers to non-polar (TiO$_2$)$^0$ layers partially occupy the Ti 3\textit{d} orbitals in the (TiO$_2$)$^0$ layers near the interfacial region, thus forming the 2DEG in the STO layer.\cite{Nakagawa_2006_NM,Thiel_2006_Sc,NazirAPL-2014} 
The third one is via a discontinuity of the spontaneous or strain-induced polarization that creates an internal electrostatic field and further produces polarization charges bound at the interface, creating the 2DEG. One typical example 2DEG system using this mechanism is the traditional semiconductor material Zn$_{1-x}$Mg$_x$O/ZnO.\cite{Tsukazaki_SC_2007,Tampo_APL_2008,Tsukazaki_namt_2008} In the Zn$_{1-x}$Mg$_x$O/ZnO HS system, the polarization-induced internal electrostatic filed produces and confines the uncompensated bound charge at the interface, forming the 2DEG.\cite{Betancourt_PRB_2015} 
In addition to these three mechanisms, several other factors such as  oxygen vacancies\cite{Chen_2011_NL, Syro_N_2011,Meevasana_nmat_2011,Ariando_2013_PRX,Zunger-nc-2014}  and cation intermixing\cite{Nakagawa_2006_NM,Kalabukhov_2007_PRB} were also proposed to be able to produce metallic states. 
For instance, it was recently revealed that tuning redox reactions on the surface of STO substrate might be one approach to produce oxygen vacancies and the resulting interfacial conductivity.\cite{Chen_2011_NL}

Compared to the success of generating 2DEG in the LAO/STO system via the polar discontinuity,\cite{Ohtomo_2004_N,Okamoto} there have been few reports on the possibility to produce the 2DEG in the perovskite oxide HS using the polarization. Very recently, Chen and co-workers prepared CaZrO$_3$/SrTiO$_3$ (CZO/STO) perovskite oxide HS by depositing the CZO film on the (001) TiO$_2$-terminated STO substrate using Pulsed Laser Deposition (PLD) and observed an insulator-to-metal transition in the HS system, when the CZO film thickness \textit{t} increases up to around 6 unit cells (uc).\cite{Chen_NL_2015} The CZO/STO HS shows metallic behaviour at \textit{t} $>$ 6 uc but   insulating at \textit{t} $\leq$ 6 uc. By combinatorial Scanning Transmission Electron Microscopy (STEM) and Electron Energy-Loss Spectroscopy (EELS) characterization, they found that the  Ca$^{2+}$ and Zr$^{4+}$ cations in the CZO film  move towards the STO substrate, which is caused by the lattice-mismatch-induced compressive strain on the CZO film.
This indicates that the relative displacement between the cations and anions in the CZO film produces a piezoelectric polarization, which plays a crucial role in forming the 2DEG in the CZO/STO HS system. 
This experimental finding, for the first time, has provided a solid proof to produce a high-mobility ($\sim$60, 000 cm$^2$ V$^{-1}$ s$^{-1}$ at 2K) 2DEG in the non-polar perovskite oxide HS using the piezoelectric polarization effects. 
In spite of the encouraging experimental finding, several open fundamental questions arise naturally such as the influence of the surface termination on the interfacial electronic properties and the dependence of the electron transport property on the thickness of the CZO film.
Therefore, from the theoretical viewpoint, it is still very necessary to deeply understand the underlying working mechanism to produce the interfacial 2DEG in the CZO/STO HS system.

In this work, to reveal the underlying mechanism to produce the interfacial conductivity in the CZO/STO HS system, we studied the structural and electronic properties of the HS system using first-principles electronic structure calculations. Four types of CZO/STO HS-based slab models, \textit{i.e.}, TiO$_2$/CaO and SrO/ZrO$_2$ interface models with CaO and ZrO$_2$ surface terminations in each model, were studied. 
We investigated the strain-induced polarization and resulting electronic properties in these models, and found that there exists an insulator-to-metal transition as the CZO film thickness increases.  
The relative displacement between the cations and anions, polarization direction and strength, charge transfer, and the critical thickness of the CZO film for forming the interfacial metallic states were elucidated. This work  provides a clear picture for the strain-induced polarization and consequent formation of the 2DEG in the nonpolar perovskite-based HS system.

\section{STRUCTURAL MODELING AND COMPUTATIONAL DETAILS}
In principle, two types of interface structures, \textit{i.e.}, TiO$_2$/CaO and SrO/ZrO$_2$, can be modeled. In each model, the CZO film is deposited on the TiO$_2$-terminated and SrO-terminated (STO)$_{10}$ substrate along the [001] direction, respectively,  in which 10 uc of the STO was used to model the substrate. Considering that different surface termination of the CZO film, \textit{i.e.}, the CaO surface and the ZrO$_2$ surface, may influence the polarization and the resulting electronic property of the HS system, we added a vacuum layer of about 15 {\AA} along the [001] direction of the HS model above the CZO film. Hence, to perform a comprehensive comparison study, four types of HS-based slab models, \textit{i.e.}, TiO$_2$/CaO interface models with CaO and ZrO$_2$ surface termination, and SrO/ZrO$_2$ interface models with CaO and ZrO$_2$ surface termination, were simulated in this work.
Another reason for doing this is based on the experimental consideration. Normally, that is, under  stoichiometric condition, the TiO$_2$/CaO interface model is terminated with ZrO$_2$ surface, and the SrO/ZrO$_2$ interface model is terminated with CaO surface. Actually, by tuning the chemical potential of the composition of the CZO film in the experiment, that is, under non-stoichiometric condition, the TiO$_2$/CaO interface model may be terminated with CaO surface, and similarly, the SrO/ZrO$_2$ interface model may be terminated with ZrO$_2$ surface. A recent experiment revealed that, in the well-known (TiO$_2$)$^0$/(LaO)$^{1+}$ LAO/STO HS system, the Sr atoms can even diffuse from the interface to the surface, forming SrO surface layer.\cite{Treske_AM_2014}

As discussed later, after structural relaxation, strong polarization may occur in the CZO/STO HS models. The periodic boundary condition in the supercell approach determines that there might exist an electric-field interaction between the surface of the STO substrate and the surface of the CZO film, though the strength strongly depends on the vacuum thickness. To exclude the influence of such type of electric-filed interaction on the polarization and electronic property, we made test calculations by inserting a 40 {\AA} vacuum layer into the TiO$_2$-terminated CZO/STO HS model, in which the electrostatic interaction between the neighboring surfaces can be negligible. Our test calculations show that the two CZO/STO HS models with a 15 {\AA} vacuum layer and a 40 {\AA} vacuum layer give exactly same results, suggesting that a vacuum of 15 {\AA} in the CZO/STO HS-based slab system is appropriate to avoid the presumably electrostatic interaction.

The STO crystal  has a cubic phase (space group no. 221, $Pm\bar{3}m$) at room temperature with a lattice parameter of 3.905.{\AA}\cite{Ohtomo_2004_N} The CZO crystallizes in an orthorhombic (space group no. 62, $Pnma$) at room temperature,\cite{Mathews_1991_JMSL} and there exists a tilting and rotation of the ZrO$_6$ octahedra (corresponding to Zr-O-Zr bending).\cite{Woodward_AC_1997} To describe the epitaxial perovskite-based heterostructure, one convenient approach is to treat the distorted perovskite as pseudo-cubic phase. In this work,  the distorted CZO crystal was modeled in a pseudo-cubic phase with a lattice constant of 4.012 {\AA},\cite{Chen_NL_2015}, which lead to a lattice mismatch of approximately $2.7\%$. To resemble the experimental epitaxial material  growth process, all the ions are fully relaxed by minimizing the atomic forces up to 0.03 eV/{\AA}, while the cell parameters along the \textit{x}- and \textit{y}-axis were fixed.

Our Density Functional Theory (DFT) electronic structure calculations were carried out using the Vienna \textit{Ab-initio} Simulation Package (VASP).\cite{VASP_PRB,VASP_CMS} The Projector Augmented Wave (PAW) potentials were used for electron-ion interactions\cite{PAW}and the Generalized Gradient Approximation (GGA) parameterized by Perdew-Burke-Ernzerhof (PBE) was used for exchange-correlation functional.\cite{PBE} The cut-off energy of 450 eV for the plane wave basis set and the $10\times10\times1$ $k$-space grid in the irreducible wedge of the Brillouin zone were used.  The self-consistency was assumed for a total energy convergence of less than 10$^{-5}$ eV. 
The density of states (DOS) was calculated using the Gaussian smearing of 0.05 eV.

\section{RESULTS AND DISCUSSION}
\subsection{TiO$_2$/CaO Interface Model} 
\subsubsection{CaO Surface Termination} 

\begin{figure}[t]
\centering
\includegraphics[width=0.45\textwidth,clip]{fig1}
\caption{(Color online) Calculated total DOS for unrelaxed (first column) and relaxed (second column) 
(CZO)$_n$/STO (\textit{n} = 3.5, 4.5, 5.5, and 6.5) HS-based slab (TiO$_2$/CaO interface model) with CaO surface termination. The vertical dashed line indicates the Fermi level in this and each subsequent DOS plot.}

\end{figure} 

\begin{figure}[t]
\includegraphics[width=0.45\textwidth,clip]{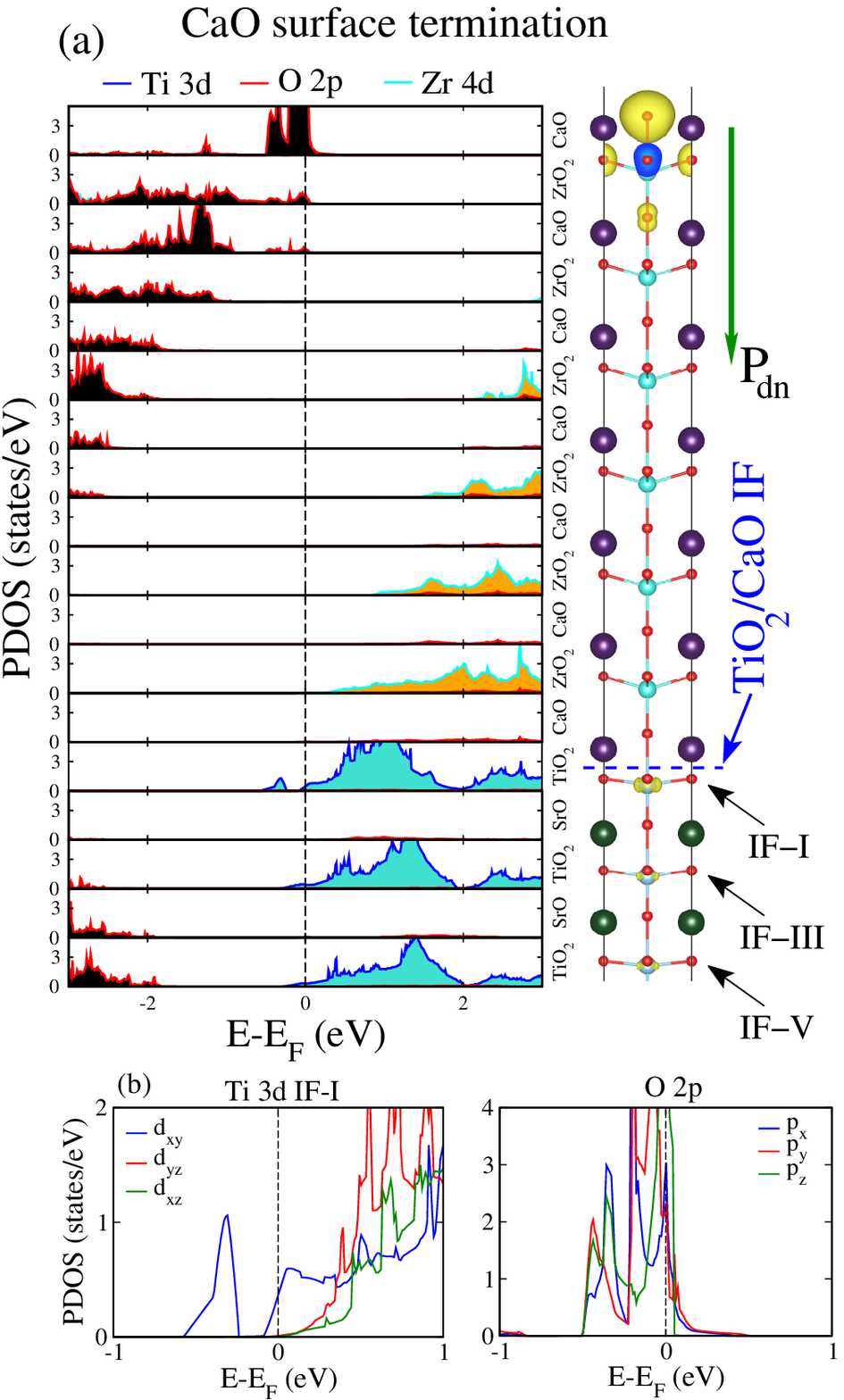}
\caption{(Color online) Calculated (a) layer-resolved partial DOS for relaxed (CZO)$_{6.5}$/STO slab (TiO$_2$/CaO interface model) with CaO surface termination, along with the charge density projected on the bands forming the metallic states near the interface and surface region. (b) Orbital-resolved partial DOS for Ti atoms at the IF-I TiO$_2$ layer and for O atom at the surface CaO layer.}
\end{figure}

We first examined the polarization and electronic property of the TiO$_2$/CaO interface model with CaO surface termination. To investigate the critical thickness of the CZO film for forming interfacial metallic states, we modeled the CZO/STO HS system by depositing the CZO film with various uc on the TiO$_2$-terminated STO substrate along the [001] direction. Hereafter, these interface systems are referred to as (CZO)$_n$/STO, in which \textit{n} denotes the number of the CZO uc. To explicitly show the polarization-induced electronic property modification, we calculated the total DOS for unrelaxed (first column) and relaxed (second column) (CZO)$_n$/STO (\textit{n} = 3.5, 4.5, 5.5, and 6.5) HS-based slab system in Fig.\ 1. 
Our calculated total DOS clearly exhibits an insulating nature for all the unrelaxed CZO/STO HS-based slab systems. This implies that there is no charge transfer between the nonpolar CZO film and STO substrate, unlike the polar/nonpolar LAO/STO HS system. 
In contrast, the relaxed CZO/STO HS models (second column in Fig.\ 1)  display an insulator-to-metal transition when the number of the CZO uc \textit{n} increases up to 6.5. For the models with the CZO uc \textit{n} less than 6.5, all the systems are insulating.  This indicates that the critical thickness of the CZO film for forming the conducting state is about 6.5 CZO uc, which is in an excellent agreement with the recent experimental finding that the CZO/STO HS model shows metallic behavior at \textit{n} $>$ 6 uc.\cite{Chen_NL_2015}

To reveal the origin of metallic states in the CZO/STO HS system, we calculated layer-resolved DOS of the (CZO)$_{6.5}$/STO HS system with  CaO surface termination in Fig.\ 2a. For a direct view of the each layer's contribution to the metallic states, we also plotted the three-dimensional (3D) charge density projected on the bands forming the metallic states near the Fermi level.  Our results show that O 2\textit{p} states in the CZO layers significantly shift toward higher energy when the CZO layer moves from the interface to the vacuum. 
At the CaO surface layer, occupied O 2\textit{p} states cross the Fermi level, producing some unoccupied states. The O 2\textit{p} states from the 2nd ZrO$_2$ and 3rd CaO surface layers also cross the Fermi level by producing minor unoccupied states.  This indicates that the CZO/STO HS forms \textit{p}-type conducting states on the surface. 
In contrast, at the interface, some Ti 3\textit{d} states cross the Fermi level by forming some occupied states, thus producing \textit{n}-type conducting states. 
For convenience of the discussion, the 1st, 3rd, and 5th layers of STO was defined as IF-I, IF-III, and IF-V, respectively. Our calculations clearly show that most of the interfacial metallic states near the Fermi level come from the IF-I TiO$_2$ layer of the STO substrate, along with a small contribution from the IF-III and IF-V TiO$_2$ layers.
Further partial DOS analysis indicates that all the other layers away from the interface exhibit an insulating behavior,  showing a typical character of the 2DEG.

The calculated charge density plot in Fig.\ 2a clearly reflects the same pattern that the conducting states mostly come from three sequential TiO$_2$ layers close to the interface, surface CaO layer, and subsurface ZrO$_2$ and CaO layers.  
The geometrical structure in the charge density plot shows that the compressive strain exerted on the CZO film leads to the downward movement of the 
Ca$^{2+}$ and Zr$^{4+}$ cations towards the interface, causing a relative displacement between the cations and anions,  
thus producing a strong ``down" polarization, $P_{dn}$. A slight downward distortion of the Ti cation away from the interface also occurs in the first few layers of the STO substrate, and thus there exists a polarization discontinuity between the CZO film and STO substrate. As a result, the strong polarization produces an internal electrical field in the CZO film, which leads to the charge transfer from the occupied O 2\textit{p} orbitals in the surface to the Ti 3\textit{d} orbitals at the interface, though the whole CZO/STO HS system keeps the charge neutrality. The transferred electrons are mainly captured by the localized Ti 3\textit{d} orbitals of the first three TiO$_2$ layer in the STO substrate, which forms \textit{n}-type interface conducting states. 
 Moreover, our calculations show that the Zr  4\textit{d} orbital substantially shifts towards the lower energy as the the ZrO$_2$ layers move from the surface to the interface (a line is marked in the STO layers for a guide to the eye), though it has no contribution to the interfacial conductivity. This implies a significant band bending between the STO substrate and the CZO film.  
  In addition, the calculated orbital-resolved partial DOS in Fig.\ 2b shows that the Ti 3\textit{d$_{xy}$} is solely responsible for the interfacial metallic states as in the case of the well-known LAO/STO HS system,\cite{You_2013_PRB,Nazir-AMI_2014} which agrees well with the \textit{d$_{xy}$}-character-like orbital shape in the charge density plot, while the \textit{d$_{yz}$} and \textit{d$_{xz}$} orbitals remain unoccupied and stay at higher energies in the conduction band.  In contrast, the orbital-resolved DOS of the surface CaO layer exhibits that all the O 2\textit{p} orbitals cross the Fermi level, forming the metallic states. In short, our theoretical calculations for the HS model with CaO surface termination provide an excellent explanation for the recent experimental 2DEG in the nonpolar CZO/STO HS system.\cite{Chen_NL_2015}

\subsubsection{ZrO$_2$ Surface Termination}

Next, we studied  TiO$_2$/CaO interface model with ZrO$_2$ surface termination. We calculated the total DOS for unrelaxed (first column) and relaxed (second column) (CZO)$_{n}$/STO (\textit{n} = 5, 6, 7, and 8) HS models in Fig.\ 3. As in the case of the HS model with CaO surface termination, all the unrelaxed HS models with ZrO$_2$ surface termination exhibit an insulating behavior (first column in Fig.\ 3). 
In contrast, the relaxed HS models with \textit{n} $\geq$ 8 show a typical conducting character, with a number of states crossing Fermi level (second column in Fig.\ 3). As discussed later, the conducting property is induced by the polarization-driven charge transfer from the middle CaO layer in the CZO film to the interfacial TiO$_2$ layers, which lead to the partially unoccupied O 2\textit{p} orbitals and the partially occupied Ti 3\textit{d} orbitals and thus \textit{p}-type and \textit{n}-type conducting states, that is, the co-existence of Two-Dimensional-Hole-Gas (2DHG) and 2DEG. For the (CZO)$_{n}$/STO HS models with \textit{n} $\in$ \{5, 6, and 7\}, 
the Fermi level is pinned close to the valence band maximum, and there exist only minor states just above the Fermi level, showing \textit{p}-type electronic structure character. The partial DOS analysis for these systems shows that the minor gap states  arise from the partially unoccupied O 2\textit{p} orbital and Zr 4\textit{d} orbital, see Fig.\ 1S in the Supporting Information. As discussed later, this is because the CZO film form a ``up" polarization domain toward the vacuum, leading to the charge transfer from the O 2\textit{p} orbital of the STO substrate to the Zr 4\textit{d} orbital of the surface ZrO$_2$ layer.
Similar electronic property also appears in the HS model with less CZO layers, \textit{i.e.}, (CZO)$_{n}$/STO (\textit{n}=3 and 4), while the HS model (CZO)$_{2}$/STO exhibits an insulating behavior, indicating an insulating-to-\textit{p}-type-conductivity transition at \textit{n}=3, see Fig.\ 2S of the Supporting Information. In the (CZO)$_{2}$/STO model, as confirmed from the structural analysis, the polarization in the CZO film is much weaker than that in other models, which is not capable of driving the charge transfer between the CZO film and STO substrate and thus the system keeps insulating. 

\begin{figure}[t]
\centering
\includegraphics[width=0.45\textwidth,clip]{fig3}
\caption{(Color online) Calculated total DO for unrelaxed (first column) and relaxed (second column) (CZO)$_n$/STO (\textit{n} = 5, 6, 7, and 8) HS-based slab (TiO$_2$/CaO interface model) with ZrO$_2$ surface termination.} 
\end{figure} 

\begin{figure}[t]
\centering
\includegraphics[width=0.40\textwidth,clip]{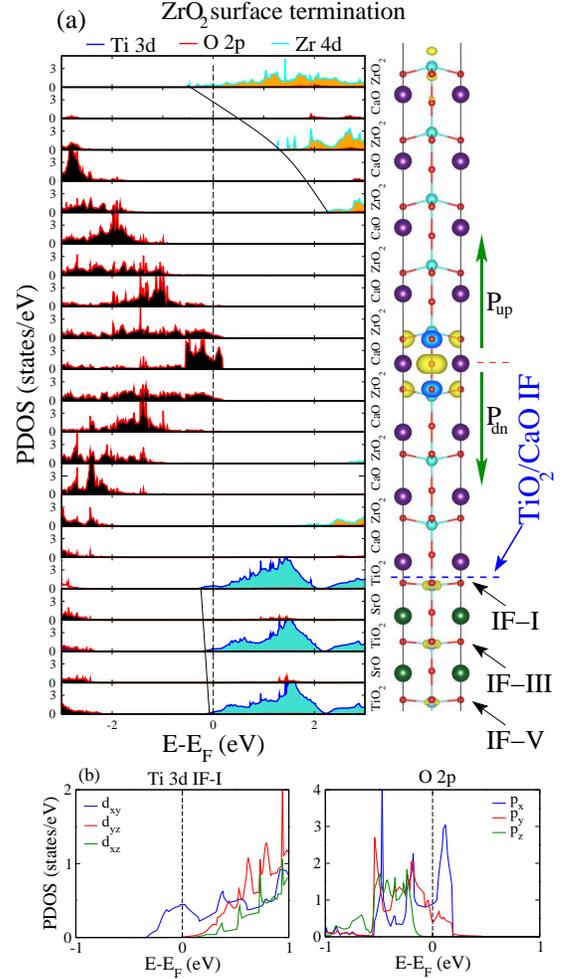}
\caption{(Color online) Calculated (a) layer-resolved partial DOS for relaxed (CZO)$_8$/STO slab (TiO$_2$/CaO interface model) with ZrO$_2$ surface termination, along with the charge density projected on the bands forming the metallic states near the interface and surface region. (b) Orbital-resolved partial DOS for Ti atoms at the IF-I TiO$_2$ layer and for O atom from the middle CaO layer.}
\end{figure}

The analysis for these relaxed HS structures indicates that the Ca and Zr cations in the CZO film move towards the vacuum, producing relative displacements between the cations and anions and the resulting ``up" polarization, $P_{up}$, which lead to the minor charge transfer from the TiO$_2$ layer to the surface ZrO$_2$ layer, see Fig.\ 3S in the Supporting Information.
It is worth emphasizing that the HS models (CZO)$_{n}$/STO (\textit{n} = 3, 4, 5, 6, and 7) do not form 2DEG because their conductivity is caused by the partially unoccupied O 2\textit{p} orbitals instead of Ti 3\textit{d} orbitals. Moreover, their resistivity is expected to be much higher than that of (CZO)$_{n}$/STO (\textit{n} $\geq$ 8) because of relatively low carrier density, as shown in the DOS plot of Fig.\ 3.

To further explore the formation mechanism of the 2DEG in the TiO$_2$/CaO interface model with ZrO$_2$ surface termination,  we calculated the layer-resolved DOS of the (CZO)$_{8}$/STO HS system in Fig.\ 4, along with the 3D charge density projected on the bands forming the metallic states near the Fermi level. One can clearly see that the metallic states mainly come from two components: the partially unoccupied O 2\textit{p} orbitals from the middle CaO layer and adjacent ZrO$_2$ layers, and the partially occupied Ti 3\textit{d} orbitals near the interfacial region in the STO substrate, along with few partially occupied Zr 4\textit{d} orbitals on the surface. Interestingly, in contrast to the model with CaO surface termination, the (CZO)$_{n}$/STO HS with \textit{n} $\geq$ 8 model with ZrO$_2$ surface termination exhibits two polarization domains with opposite directions, as shown in the charge density plot and Fig.\ 3S in the Supporting Information. Above the middle CaO layer, the  Ca$^{2+}$ and Zr$^{4+}$ cations in the CZO film move towards the vacuum, leading to a ``up" polarization, $P_{up}$. This strong polarization induces a charge transfer from the occupied O 2\textit{p} orbitals of the middle CaO layer to the surface ZrO$_2$ layer, producing \textit{n}-type conductivity on the surface ZrO$_2$ layer and \textit{p}-type conductivity in the middle CaO layer, see Fig.\ 4a and Fig.\ 3S of the Supporting Information. In contrast, below the middle CaO layer, the cations of the CZO film move towards the STO substrate, and the relative displacement between the cations and anions produces a ``down" polarization $P_{dn}$ that causes the charge transfer from the CZO film to the STO substrate. It is noted that, similar to the case of the HS model with CaO surface termination, the STO substrate exhibit a much weaker distortion within the first three TiO$_2$ layers than the CZO film, and the deep TiO$_2$ layers nearly show no distortion. This implies that  a polarization discontinuity  exists between the CZO film and STO substrate. Therefore, the transferred charge from the middle CaO layer to the STO substrate was captured by the localized Ti 3\textit{d} orbitals in the first few interfacial TiO$_2$ layers of the STO substrate, forming the interfacial metallic states. The partial DOS analysis displays that only the first three TiO$_2$ layers, \textit{i.e.}, IF-I, IF-III, and IF-V layers, contribute to interfacial conducting states, which confirms the formation of the 2DEG. The calculated orbital-resolved partial DOS plot shows that the interfacial metallic states on the IF-I layer only come from the $d_{xy}$ orbitals of Ti ions, see Fig.\ 4b. 
It is particularly worth mentioning that, in the CZO/STO HS system with ZrO$_2$ surface termination, besides the 2DEG in the interfacial region, there also exists a 2DHG in the middle layer of the CZO film, which is contributed by the $p_x$/$p_y$  orbitals of O ions. This is because the ``down" and ``up" polarization in the CZO film leads to the charge transfer from CZO film to the STO substrate and the surface ZrO$_2$ layer simultaneously,  thus producing the \textit{p}-type hole states in the middle CaO and adjacent ZrO$_2$ layers.

\subsection{SrO/ZrO$_2$ Interface Model} 
\subsubsection{CaO Surface Termination}

\begin{figure}[t]
\centering
\includegraphics[width=0.45\textwidth,clip]{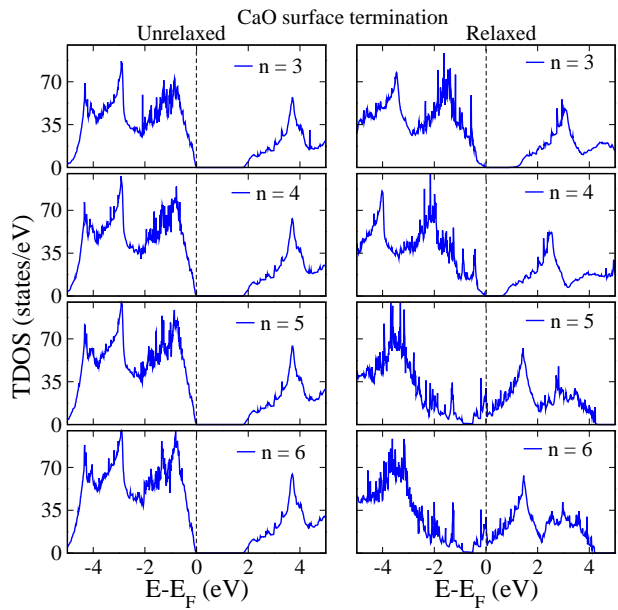}
\caption{(Color online) Calculated total DOS for unrelaxed (first column) and relaxed (second column) (CZO)$_n$/STO (\textit{n} = 3, 4, 5, and 6) HS-based slab (SrO/ZrO$_2$ interface model) with CaO surface termination.} 
\end{figure}

\begin{figure}[t]
\centering
\includegraphics[width=0.45\textwidth,clip]{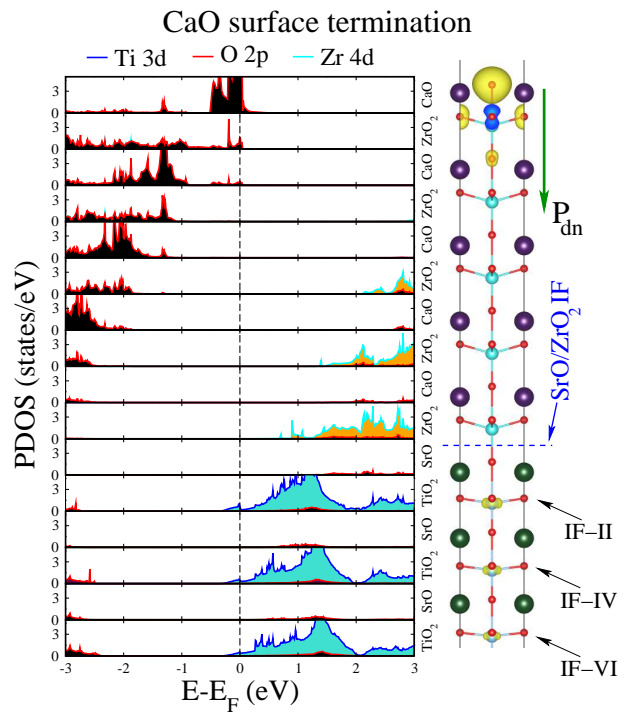}
\caption{(Color online) Calculated layer-resolved partial DOS for relaxed (CZO)$_5$/STO slab  (SrO/ZrO$_2$ interface model) with CaO surface termination, along with the charge density projected on the bands forming the metallic states near the interface and surface region.}
\end{figure}

In this section, we turn to the case of SrO/ZrO$_2$ interface model by beginning with the HS model with CaO surface termination. 
To explore the possible critical thickness for forming interfacial metallic states, we performed calculations for (CZO)$_n$/STO (\textit{n} = 3, 4, 5, and 6) HS model. The calculated total DOS for the unrelaxed and relaxed  HS systems are shown in Fig.\ 5. As expected, before structural relaxation, all the unrelaxed HS systems exhibit insulating character because there is no polarization in the CZO film. After structural relaxation, the (CZO)$_n$/STO models with \textit{n} $\in$ \{3 and 4\} exhibit insulating behavior, and as \textit{n} increases up to 5, an insulator-to-metal transition occurs. This indicates that, for the SrO/ZrO$_2$ interface model with the CaO surface termination, 5 CZO uc is the critical thickness for forming conductivity, which is less than that of the TiO$_2$/CaO interface model with the same surface termination.  

To investigate the origin of the metallic states in the SrO/ZrO$_2$ interface model, we calculated its layer-resolved DOS and 3D charge density plot projected on the bands forming the metallic states, see Fig.\ 6.   Similar to the case of TiO$_2$/CaO interface model, the compressive strain applied in the CZO film causes a downward movement of the Ca$^{2+}$ and Zr$^{4+}$ cations to the interface, leading to a ``down" polarization, $P_{dn}$. 
This $P_{dn}$ leads to a charge transfer from the occupied O 2\textit{p} orbitals in the surface CaO layer to the Ti 3\textit{d} orbitals at the interface. The transferred electrons are mainly captured by the localized Ti 3\textit{d} orbitals of the first three TiO$_2$ layer in the STO substrate, producing \textit{n}-type interface conducting states.

\begin{figure}[t]
\centering
\includegraphics[width=0.45\textwidth,clip]{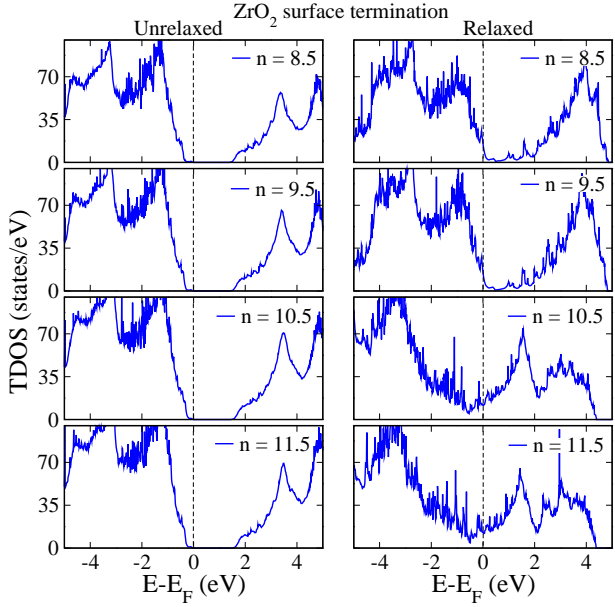}
\caption{(Color online) Calculated total DOS at for unrelaxed (first column) and relaxed (second column) (CZO)$_n$/STO (\textit{n} = 8.5, 9.5, 10.5, and 11.5) HS-based slab (SrO/ZrO$_2$ interface model) with ZrO$_2$ surface termination.} 
\end{figure}

\begin{figure}[t]
\centering
\includegraphics[width=0.37\textwidth,clip]{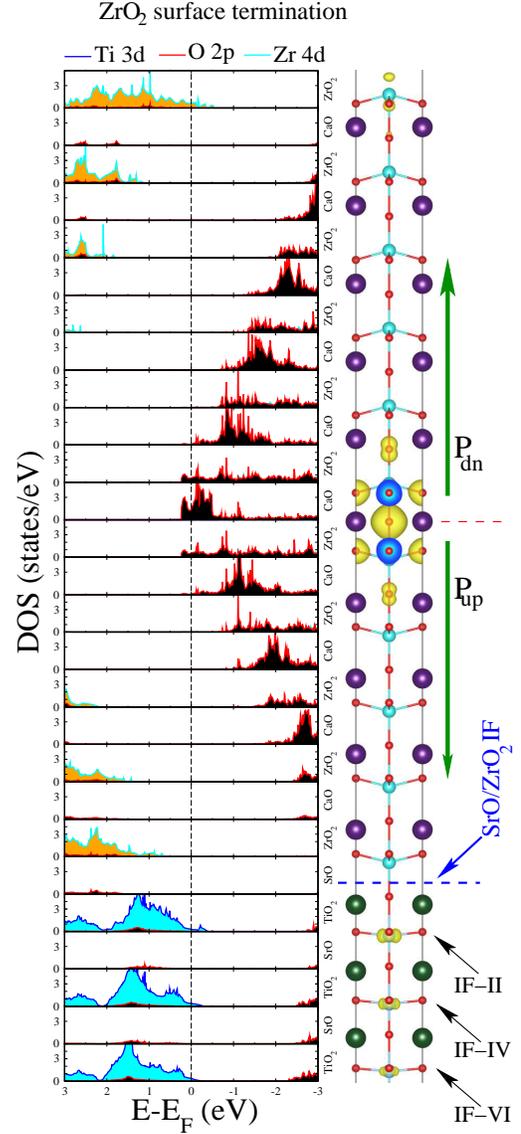}
\caption{(Color online) Calculated layer-resolved partial DOS for relaxed (CZO)$_{10.5}$/STO slab  (SrO/ZrO$_2$ interface model) with ZrO$_2$ surface termination, along with the charge density projected on the bands forming the metallic states near the interface and surface region.}
\end{figure}

\subsubsection{ZrO$_2$ Surface Termination}
Next, we studied the SrO/ZrO$_2$ interface model with ZrO$_2$ surface termination. The calculated total DOS for the unrelaxed and relaxed (CZO)$_n$/STO (\textit{n} = 8.5, 9.5. 10.5, and 11.5) HS model is shown in Fig.\ 7. Before the structural relaxation, all the models show an insulating property. After structural relaxation, the CZO/STO HS models with \textit{n} = 8.5 and 9.5 show a \textit{p}-type conductivity, in which the Fermi level is pinned slightly below the valence band maximum with minor states crossing the Fermi level. Similar electronic structure character is also found in the  (CZO)$_n$/STO HS models with \textit{n} $\in$ \{4.5, 5.5, 6.5, and 7.5\}. The partial DOS analysis informs that the minor states above the Fermi level in these HS models come from the unoccupied O 2\textit{p} orbital in the STO substrate. This is because, in these relaxed (CZO)$_n$/STO HS models (\textit{n} = 4.5 to 9.5), the``up" polarization, $P_{up}$, caused by the relative displacements between the cations and anions, leads to a charge transfer from the STO substrate to the surface ZrO$_2$ layer (see Fig.\ 4S in the Supporting Information), similar to the case of the TiO$_2$/CaO interface model with ZrO$_2$ surface termination (Fig.\ 3S in the Supporting Information).

At $n$ $\geq$ 10.5, the (CZO)$_n$/STO HS system exhibits metallic states. The calculated  layer-resolved DOS of the (CZO)$_{10.5}$/STO HS model and 3D charge density projected on the bands forming the metallic states are shown in Fig.\ 8. It shows that two polarization domains with opposite directions P$_{up}$ and P$_{dn}$  in the CZO film, separated by the middle CZO layer, are formed in the CZO film. In the $P_{up}$ domain, the Ca$^{2+}$ and Zr$^{4+}$ cations move towards the vacuum, and the polarization induces a charge  transfer from the middle CaO layer to the surface ZrO$_2$ layer, creating \textit{n}-type conductivity on the surface ZrO$_2$ layer and \textit{p}-type conductivity in the middle CaO layer, see Fig.\ 8. In the $P_{dn}$ domain, in contrast, the cations of the CZO film move towards the STO substrate, and the $P_{dn}$ polarization leads to a charge transfer from the middle CaO layer to the STO substrate, producing \textit{n}-type interfacial conductivity. 
This indicates that, in the SrO/ZrO$_2$ interface model with ZrO$_2$ surface termination, the two polarization domains lead to the co-existence of the 2DEG on the interface and the 2DHG in the CZO film. 
The critical thickness of the CZO film for forming such an electronic state is about 10.5 uc. 
In short, our results indicate that over all, the polarization and electronic structure characters of the SrO/ZrO$_2$ interface model are similar to those of the TiO$_2$/CaO interface model for the each type of the surface termination, though the required critical thickness to form the interfacial conducting states in the STO substrate is different. 
Hence, it is expected that the 2DEG can also be observed in the SrO/ZrO$_2$ interface model, as in the case of the TiO$_2$/CaO interface model.\cite{Chen_NL_2015}


\begin{figure}[t]
\includegraphics[width=0.45\textwidth,clip]{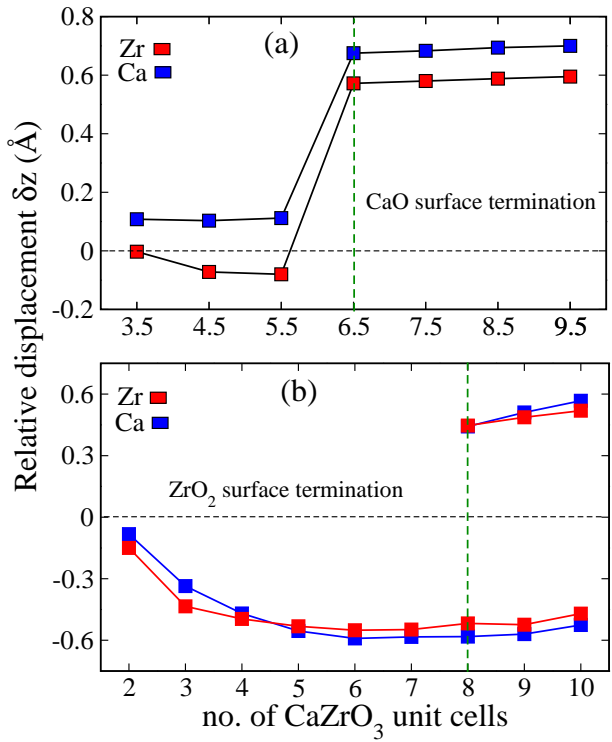}
\caption{(Color online) Calculated average relative displacement ($\delta z$) of Ca and Zr cations with respect to O ions in the corresponding CaO and ZrO$_2$ planes in the CZO film for the TiO$_2$/CaO interface model
with (a) CaO and (b) ZrO$_2$ surface termination. 
Vertical dashed green lines indicate the critical thickness of the CZO films for forming 2DEG at the interface in each model. For the (CZO)$_n$/STO HS model with ZrO$_2$ surface termination, at \textit{n}$\geq$ 8, two polarization domains, $P_{up}$ and $P_{dn}$, are formed, and the Ca/Zr cations move in opposite directions in the two domains.}
\end{figure}

\subsection{Structural Relaxation}

Since the TiO$_2$/CaO interface model shares similar polarization and electronic structure characters with those of the SrO/ZrO$_2$ interface model, in the following discussions, we only take the TiO$_2$/CaO interface model as an example to explore the polarization property and electron carrier density for CZO/STO HS system. As discussed above, the structural optimization plays a crucial role in producing the metallic states in all the CZO/STO HS models.  Specifically, the strain-induced relative displacement between the cations and anions, \textit{i.e.}, polarization, leads to the formation of 2DEG (2DHG) in the CZO/STO HS system. To quantitatively characterize the relationship between the polarization and the polarization-induced 2DEG (2DHG) in the two CZO/STO HS models, we plotted the average relative displacement ($\delta z$) of the Ca/Zr cations in the CZO film along the \textit{c}-axis as a function of the number of the CZO uc (\textit{n}) in Fig.\ 9 in each case. The $\delta z$ of Ca/Zr cations was calculated with respect to the  oxygen ions in the corresponding CaO and ZrO$_2$ plane. The  positive and negative values indicate that Ca/Zr cations move towards and away from the interface, respectively. Our results clearly show that, in the HS model with CaO surface termination, $\delta z_{Ca}$ $>$ 0 for the whole range of \textit{n}, indicating that Ca cations move towards the interface, while Zr cations slightly move away from the interface at \textit{n} $<$ 6.5, see Fig.\ 9a. This implies that the total polarization in the CZO film is almost negligible below the critical thickness. 
When \textit{n} increases up to 6.5, \textit{i.e.}, the critical thickness of the CZO film for forming the 2DEG, $\delta z$ is about 0.65 {\AA}, and as \textit{n} further increases, $\delta z$ slightly increases.
For the (CZO)$_{n}$/STO HS model with ZrO$_2$ surface termination, as mentioned above, at \textit{n} $\leq$ 7, the Ca/Zr cations move  away from the interface, \textit{i.e.},  $\delta z$ $<$ 0. As shown in Fig.\ 9b, when \textit{n} increases from 2 to 7, $\mid\delta$z $\mid$ gradually increases up to 0.5 {\AA}, and as \textit{n} further increases, $\mid\delta$z $\mid$ nearly keeps constant. At \textit{n} $\geq$ 8, two polarization domains, $P_{up}$ and $P_{dn}$, are formed, as shown in Fig.\ 4a. That is to say, the Ca/Zr cations move from the middle CZO layers towards the vacuum and STO substrate, respectively, in the $P_{up}$ and $P_{dn}$ polarization domain, and $\mid\delta$z $\mid$ in the two polarization domains is comparable.

To quantitatively characterize the polarization strength, we calculated the polarization \textit{P} of the CZO film for the two HS models using the following formula:\cite{Vanderbilt_PRL_1994, Yun_2010_JAP}

\begin{equation}
\centering
P = \frac{e}{\Omega}\sum\limits_{i=1}^N Z_i^*\delta z_i  
\label{eqn}
\end{equation}

where $\Omega$ is the volume of the CZO film, \textit{N} is the number of cation in the CZO film, $Z_i^*$ is Born effective charge for each cation (Ca/Zr), and $\delta z_i$ is the relative displacement between the \textit{i}th cation (Ca/Zr) and the anion (O) in the corresponding CaO/ZrO$_2$ plane, \textit{i.e.}, $\delta z_i$=$z_O$-$z_{Ca/Zr}$.
Our calculated Born effective charges $Z_i^*$ are 2.27 and 4.93 for Ca and Zr, respectively, and $-$3.88 and $-$1.66 for O ions in the CaO and ZrO$_2$ planes, respectively, for the tetragonal bulk CZO. For the CZO/STO HS model with CaO surface termination, at \textit{n}=6.5, \textit{i.e.}, critical thickness of forming 2DEG, the estimated polarization  is about 66.5 $\mu$C/cm$^2$. For the HS model with ZrO$_2$ surface termination, at \textit{n}=8, two polarization domains are formed in the CZO film. The polarization of each domain was calculated as about  57.4 $\mu$C/cm$^2$ for $P_{dn}$ and 46.4 $\mu$C/cm$^2$ for $P_{up}$, respectively. These values are comparable with the the closely related SrZrO$_3$/STO superlattice system, in which the polarization was estimated to be around 41-43 $\mu$C/cm$^2$.\cite{Kun_PRB_2007, Yang_SSC_2006} Nevertheless, it is noted that the polarization value calculated from first-principles calculation is much larger than the experimental value of 3.5 $\mu$C/cm$^2$.\cite{Chen_NL_2015} This is probably because that, in the epitaxial materials growth process, as the thickness of the CZO film increases, the CZO film undergoes less compressive strain from the STO substrate and tends to recover its bulk equilibrium condition,  and thus the relative displacement $\delta z$ of the cations in the CZO film decreases significantly. In contrast, in the computational simulation, the lattice constant  of the whole HS model in the \textit{ab}-plane was fixed due to the periodicity requirement in the plane-wave-based DFT calculations, which means that the whole CZO film undergoes the same compressive strain due to the lattice mismatch, thus leading to the larger average $\delta z$ than the experimental value. Hence, this discrepancy is expected to become larger as the CZO film grows thicker.

In short, our calculations show that, when the CZO film exceeds the critical thickness, \textit{i.e.}, 6.5 uc. for the model with CaO surface termination and 8 uc. for the model with ZrO$_2$ surface termination, both the two HS systems can lead to polarization, $P_{dn}$,  and the resulting 2DEG on the interface. The difference is that only one polarization domain, $P_{dn}$, is formed in the model with CaO surface termination, while two polarization domains, \textit{i.e.}, $P_{up}$ and $P_{dn}$, are formed in the model with ZrO$_2$ surface termination. This is probably because the CaO-terminated surface and the interface share the same role in driving the $P_{dn}$ towards the STO substrate, while the ZrO$_2$-terminated surface and the interface have a competing role in driving the polarization due to the electrostatic potential discontinuity. In addition, the CZO film with CaO surface termination was predicted to be energetically more stable than that with  ZrO$_2$ surface termination,\cite{Brik_2013_SC} implying that the experimental CZO/STO HS sample is likely to be terminated with CaO surface. In fact, as discussed above, our predicted 2DEG in the HS-based slab model with CaO surface termination is more consistent with the Chen \textit{et al.}'s experimental observation.\cite{Chen_NL_2015}

\begin{figure}[t]
\centering
\includegraphics[width=0.45\textwidth,clip]{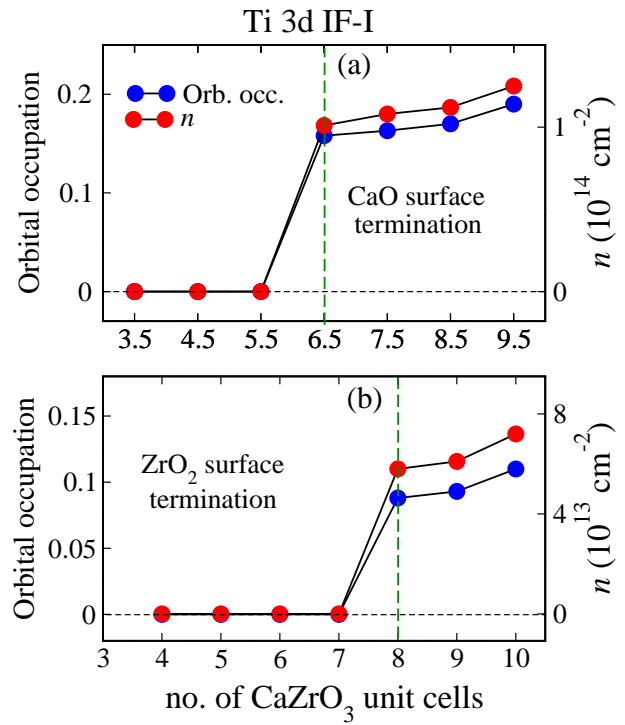}
\caption{(Color online) Calculated orbital occupation and electron carrier density of Ti 3\textit{d} orbitals from the IF-I TiO$_2$ layer in the STO substrate with respect to the number of CZO unit cells (\textit{n}) for the TiO$_2$/CaO interface model with (a) CaO and (b) ZrO$_2$ surface termination. 
Vertical dashed green lines indicate the critical thickness of the CZO films for forming 2DEG at the interface in each model. }
\end{figure}

\subsection{Interfacial Electron Carrier Density}

To examine the influence of the CZO film thickness on the interfacial charge carrier density, we computed the occupation number of the Ti 3\textit{d} orbitals from the interfacial (IF-I) TiO$_2$ layer by integrating the partial DOS of the Ti 3\textit{d} occupied states near the Fermi level as well as the interfacial charge carrier density for the TiO$_2$/CaO interface model with CaO and ZrO$_2$ surface terminations in Fig.\ 10. This is because that the IF-I TiO$_2$ layer plays a major role in the interfacial conductivity in the CZO/STO HS system. For the (CZO)$_n$/STO HS model (\textit{n} = 3.5, 4.5, 5.5, 6.5, 7.5, 8.5, and 9.5) with CaO surface termination, the calculated orbital occupation of Ti 3\textit{d} orbital and interfacial charge carrier density as a function of the number of CZO uc (\textit{n}) was plotted in Fig.\ 10a. Similarly, for the (CZO)$_n$/STO HS model (\textit{n} = 4, 5, 6, 7, 8, 9, and 10) with ZrO$_2$ surface termination, the corresponding orbital occupation number and interfacial charge carrier density are shown in Fig.\ 10b. As discussed above, below the critical thickness of the CZO film for forming the 2DEG at the interface, there is no electrons captured by the Ti 3\textit{d} orbitals in the IF-I TiO$_2$ layer, and thus the systems exhibit zero orbital occupation and interfacial electron carrier density.  Above the critical thickness of the CZO film, the interface becomes conductive, and as the CZO unit cell \textit{n} increases, their interfacial electron carrier density increases., which  is in a good agreement with the recent experimental finding.\cite{Chen_NL_2015}

\section{CONCLUSION}

In summary, we studied the formation of 2DEG in the CZO/STO HS-based slab systems using first-principles electronic structure calculations. Four types of CZO/STO HS-based slab systems, including TiO$_2$/CaO and SrO/ZrO$_2$ interface models with CaO and ZrO$_2$ surface terminations in each model, were modelled. It was found that, as the thickness of CZO film increases, the compressive strain exerted on the CZO film, caused by the lattice-mismatch between the CZO film and STO substrate, leads to a strong polarization in the CZO layer and the resulting interfacial 2DEG at the interface. Nevertheless, the polarization direction and critical thickness of the CZO film for forming the 2DEG strongly depend on the surface termination of the CZO film. For the HS model with CaO surface termination, a critical thickness of CZO film of approximately 6.5 and 5 unit cells is necessary to produce a strong polarization towards the STO substrate in the TiO$_2$/CaO  and SrO/ZrO$_2$ interface models, respectively. The polarization induces the charge transfer from the surface CaO layer to the first few TiO$_2$ layers in the STO substrate, thus forming the 2DEG at the interface. For the HS model with ZrO$_2$ surface termination, in contrast, as the CZO film increases up to 8 unit cells for TiO$_2$/CaO interface model and 10.5 unit cells for SrO/ZrO$_2$ interface model, the lattice-mismatch-induced strain produces two polarization domains with opposite directions in the CZO film, which points towards the interface and surface from the middle CZO layer, respectively.
The two polarization domains leads to the co-existence of the 2DEG at the interface and the 2DHG at the middle CZO film, along with the  \textit{n}-type few surface metallic states. Our results reveal the mechanism of producing the high-mobility 2DEG (2DHG) in the nonpolar CZO/STO HS system based on the stain-induced polarization discontinuity, which provides a new avenue to produce 2DEG (2DHG) in the perovskite-based HS systems.

\section{Acknowledgment}

KY thanks Dr. Jayakanth Ravichandran, Dr. Lin Xie, and Dr. Jian Luo for useful discussions. This work was supported by the start-up funds from the University of California San Diego.  


%

\end{document}